\documentclass[a4paper,fleqn]{cas-sc}

\usepackage[authoryear]{natbib}

\def\tsc#1{\csdef{#1}{\textsc{\lowercase{#1}}\xspace}}
\tsc{WGM}
\tsc{QE}
\tsc{EP}
\tsc{PMS}
\tsc{BEC}
\tsc{DE}

\usepackage[caption = false]{subfig}
\usepackage{graphicx}
\usepackage{dsfont}
\usepackage{enumitem}
\usepackage{multirow,booktabs,array}

\usepackage{soul}
\usepackage{color}
\usepackage{comment}

\usepackage{tikz}
\usetikzlibrary{decorations.pathreplacing}
\usepackage{verbatim}
\usepackage{pgfplots}
\pgfplotsset{compat=newest}

\usepackage{nomencl} 
\makenomenclature
\renewcommand*\nompreamble{\begin{multicols}{2}}
\renewcommand*\nompostamble{\end{multicols}}

\usepackage{url}
\usepackage[capitalise]{cleveref}
\Crefname{equation}{}{}

\definecolor{myGray1}{RGB}{0,0,0}
\definecolor{myGray2}{RGB}{100,100,100}
\definecolor{myGray3}{RGB}{128,128,128}
\definecolor{myGray4}{RGB}{191,191,191}
\definecolor{myGray5}{RGB}{220,220,220}

\begin{document}
\let\WriteBookmarks\relax
\def\floatpagepagefraction{1}
\def\textpagefraction{.001}
\shorttitle{A bottom-up quantification of flexibility potential from the thermal energy storage in electric space heating}
\shortauthors{L. Herre et~al.}

\title [mode = title]{A bottom-up quantification of flexibility potential from the thermal energy storage in electric space heating}                 
\tnotemark[1]

\tnotetext[1]{This document is the result of a research project funded by the Swedish Energy Agency.}


\author[1,3]{Lars Herre}[type=editor,
                        auid=000,bioid=1,
                        orcid=0000-0003-0685-0199]
\cormark[1]
\ead{lfihe@elektro.dtu.dk}
\ead[url]{https://orbit.dtu.dk/en/persons/lars-finn-herre}
\credit{Conceptualization of this study, Methodology, Software, Writing - Original draft preparation.}

\author[2]{Behrouz Nourozi}
\credit{Methodology, Software, Writing.}

\author[3]{Mohammad Reza Hesamzadeh}
\credit{Writing, Revision of draft.}

\author[2]{Qian Wang}
\credit{Revision of draft}

\author[3]{Lennart S{\"o}der}
\credit{Revision of draft}

\address[1]{DTU Denmark Technical University, Department of Electrical Engineering, 2800 Kgs. Lyngby, Denmark}
\address[2]{KTH Royal Institute of Technology, Department of Civil and Architectural Engineering, 10044 Stockholm, Sweden}
\address[3]{KTH Royal Institute of Technology, School of Electrical Engineering and Computer Science, 10044 Stockholm, Sweden}

\cortext[cor1]{Corresponding author}


\begin{abstract}
Non-generating resources such as thermostatically controlled loads (TCLs) can arbitrage energy prices and provide balancing reserves when aggregated due to their thermal energy storage capacity. Based on a performed survey of Swedish single- and two-family dwellings with electric heating, this paper quantifies the potential of TCLs to provide reserves to the power system in Sweden. To this end, dwellings with heat pumps and direct electric heaters are modeled as thermal energy storage equivalents that can be included in a linear two-stage problem formulation. We approach the operational flexibility of the TCLs by modeling a risk-averse aggregator that controls decentralized TCLs and aims to maximize its own profit. The results show a potential of 2\,${\text{GW}} / {\text{0.1\,Hz}}$ averaged over a year, and up to 6.4\,${\text{GW}}/{\text{0.1\,Hz}}$ peak capacity. Based on a sensitivity analysis we derive policy implications regarding market timing and activation signal.
\end{abstract}


\maketitle

\section*{Nomenclature}
The nomenclature is stated below. Uppercase as well as Greek letters denote input parameters, calligraphic letters denote sets and distributions, while lowercase letters are used to represent indexes and decision variables.
\addcontentsline{toc}{section}{Nomenclature}

\paragraph*{Indexes and Sets}
\begin{description} [leftmargin=4.5em,style=nextline]
    \item[$a \in \mathcal{A}$] price area / zone
    \item[$b \in \mathcal{B}$] building type, size, insulation 
    \item[$h \in \mathcal{H}$] heating type 
    \item[$k$] county
    \item[$t \in \mathcal{T}$] interval in optimization horizon
    \item[$z \in \mathcal{Z}$] climate zone 
	\item[$\tau$] current market interval
	\item[$\omega \in \Omega$] scenarios of uncertain market prices
\end{description}

\paragraph*{Individual TCL Parameters}
\begin{description} [leftmargin=4.5em,style=nextline]
    \item [$\theta_{t,b}^{\text{in}}$] indoor temperature
    \item [$\overline{\theta}_{b,h}$] maximum indoor temperature (upper deadband)
    \item [$\underline{\theta}_{b,h}$] minimum indoor temperature (lower deadband)
    \item [$R_{b}$] thermal resistance
    \item [$C_{b}$] thermal capacitance
    \item [$\eta_{h}$] thermal efficiency 
    \item [$P^{el}_{b,h}$] electric power rating
    \item [$T^{ON}_{a,b,h,t}$] time a TCL takes in powered mode from one end of the deadband to the other
    \item [$T^{OFF}_{a,b,h,t}$] time a TCL takes in unpowered mode from one end of the deadband to the other
    \item [$D_{a,b,h,t}$] duty cycle
    \item [$\hat{v}_{a,b,h,t}$] availability (binary parameter)
    \item [$\tilde{v}_{a,b,h,t}$] ability (binary parameter)
\end{description}

\paragraph*{Market Price Parameters}
\begin{description} [leftmargin=4.5em,style=nextline]
    \item[$T_c$] contract period [h]
	\item[$\Delta t$] market interval duration [h]
	\item[$\tau_h$] index of the last interval in the optimization horizon
	\item[$\lambda_{t}^{R}$] marginal price in the reserve market in interval $t$, scenario $\omega$ [$\frac{\text{\textdollar}}{\text{MWh}}$]
	\item[$\lambda_{a,t}^{DA}$] marginal price in the energy market in interval $t$, scenario $\omega$ [$\frac{\text{\textdollar}}{\text{MWh}}$]
	\item[$\lambda_{a,t,\omega}^{RT}$] real-time balancing price for real-time energy deviation in interval $t$, scenario $\omega$ [$\frac{\text{\textdollar}}{\text{MWh}}$]
	\item[$\lambda^{I}$] imbalance charge [$\frac{\text{\textdollar}}{\text{MWh}}$]
	\item[$\lambda^P$] cost associated to SOC deviation $d_\omega$ [$\frac{\text{\textdollar}}{\text{MWh}}$]
\end{description}
	
\paragraph*{TCL Parameters}
\begin{description} [leftmargin=4.5em,style=nextline]
	\item[$P^B_{a,t}$] aggregate baseline power consumption at time $t$ [MW]
	\item[$G_{a,t}$] parameter corresponding to 50\% state of energy in the thermal energy storage model at time $t$ [MWh]
	\item[$\overline{S}_{a,t},\underline{S}_{a,t}$] bounds on the state of charge $s_{a,t,\omega}$ at time $t$ [MWh]
	\item[$\overline{P}_{a}^\text{inst}$] installed power capacity in area $a$ [MW]
    \item[$\overline{P}_{a,t},\underline{P}_{a,t}$] bounds on the aggregate power consumption $p_{a,t,\omega}^{E}$ at time $t$ [MW]
	\item[$\theta_{a,t}$] ambient temperature in interval $t$ [$^o$C]
\end{description}
	
\paragraph*{Chance-Constraint, Scenario, and Risk Parameters}
\begin{description} [leftmargin=4.5em,style=nextline]
	\item[$\alpha$] risk confidence level
	\item[$\beta$] risk-aversion weight
	\item[$1-\epsilon_i$] confidence level of chance-constraint $i$
	\item[$\pi_\omega$] probability of scenario $\omega$
\end{description}
	
\paragraph*{Decision Variables}
\begin{description} [leftmargin=4.5em,style=nextline]
	\item[$d_{a,\omega}$] deviation of SOC from 50\% at $\tau_h$ in scenario $\omega$ [MWh]
	\item[$p_{a,t}^{R}$] aggregate capacity offered in the reserve market in interval $t$ [MW]
	\item[$p_{a,t,\omega}^{RT}$] aggregate power consumption in interval $t$, scenario $\omega$ [MW]
	\item[$e_{a,t,\omega}^{R}$] aggregate energy activated in the reserve market in interval $t$ [MW]
	\item[$e_{a,t}^{DA}$] aggregate day-ahead energy bid in interval $t$ [MW]
	\item[$s_{a,t,\omega}$] aggregate state of energy (SOE) at end of interval $t$, scenario $\omega$ [MWh]
	\item[$c_{a,\omega}$] profit in scenario $\omega$ [\textdollar]
 	\item[$y_a$] auxiliary risk variable [\textdollar]
 	\item[$v_{a,\omega}$] auxiliary risk variable in scenario $\omega$ [\textdollar]
 	\item[CVaR$_{a}^\alpha$] Conditional Value at Risk (CVaR) at $\alpha$ [\textdollar]
\end{description}

\paragraph*{Heat Pump Design Parameters}
\begin{description} [leftmargin=4.5em,style=nextline]
    \item [$E_z^\text{HV} $] energy for heating and ventilation in climate zone $z$ [$\frac{\text{kWh}}{\text{m}^2\,\text{year}}$]
    \item [$Q_{h,k}^\text{HV}$] work for heating and ventilation in county $k$ [$\frac{\text{MWh}}{\text{year}}$]
    \item [$\theta_k^\text{des}$] design (winter) ambient temperature in county $k$ [$^\text{o}C$]
    \item [$\theta_k^\text{DD}$] heating degree days in county $k$ [$^\text{o}C$]
    \item [$A^\text{SFD}$] surface area per dwelling in county $k$ [$\text{m}^2$]
\end{description}

\section{Introduction}
	The increasing share of variable renewable energy sources requires more flexible resources that can respond in real-time to supply/demand imbalance. 
	Demand response can be an effective means for power system operators to compensate for fluctuating renewable generation, avoid grid congestion, and cope with other contingencies. Buildings equipped with electric space heating systems can provide demand response services because their electricity consumption is flexible due to their inherently thermal inertia.
	To this end, thermostatically controlled loads (TCLs) have been proposed for energy arbitrage \citep{Mathieu2015} and providing ancillary services \citep{Callaway2009}.   
	By exploiting the inherent thermal inertia of TCLs, their electricity consumption can be varied while still meeting the desired service quality, i.e., temperature range set by the end user. 
	
	TCLs cover both	heating and cooling emission systems and energy supply systems.
	These devices include, e.g., air-conditioning (AC), chillers, electrically heated terminal units and heat pumps.
	In climates with high cooling demand like, e.g., California, AC is widely used \citep{MathieuCalifornia2015}, while in places with high heating demand like, e.g., Scandinavia, electric space heating offers the largest potential \citep{Mathieu2014dk,Nyholm2016}. 
	Within Europe, Sweden has the highest installations of heat pumps \citep{Campillo2012} despite its small population. In 2018, approximately 54\,\% of the two million single-family dwellings (SFD) in Sweden are electrically heated by different methods, including heat pumps \citep{Nilsson2019}. 
	
	In the residential sector, more than half of the energy use goes to heating and the preparations of domestic hot waters (DHW) \citep{EnergySE2020}. The number of heat pumps in Sweden is increasing and the trend shows that electric heating, heat pumps, and district heating are replacing conventional fossil based heating methods \citep{EnergySE2020}. This change towards more efficient heat pumps is driven by increasing electricity prices \citep{EnergySE2020}.
	In 2018, 17\,\% of the total Swedish electricity demand (126\,TWh) was used for heating, of which 12\,\% went to SFDs \citep{EnergySE2020}.
	Heating, ventilation and air conditioning (HVAC) for residential and commercial buildings requires a substantial share of electric energy, and ultimately drives summer peak demand in the United States \citep{Chassin2016} and the winter peak in Europe.
	The building sector is expected to play an important role in providing ancillary services to relieve stress and reduce the needs of investment for power systems \citep{Wang2019}. 
	The high share of electric heating and cooling in the total electricity demand may therefore open a large source of flexibility for the power system, if exploited.

    Simply reducing peak power demand in a building can reduce electricity expenses for the building owner and contribute to the efficiency and reliability of the electrical power grid. A control strategy for peak power reduction is proposed by \cite{Winstead2020} and examined in a simulation with 80 air-conditioning units and 40 refrigeration units.
    However, active demand response (DR) can contribute to a more cost-efficient operation of, and investment in, the electric power system as it may provide the needed flexibility to cope with the intermittent character of renewable energy sources. TCLs allow to modify their electrical load pattern without affecting the final, thermal energy service they deliver due to the thermal energy storage in the system. A large body of literature has been dedicated to the optimization and control of TCLs from various perspectives.
    
    Active Demand Response with TCLs is studied in \cite{Patteeuw2015,Arteconi2016}.
    \cite{Patteeuw2015} develop an integrated system model of the electric power system, including heat pumps and auxiliary resistance heaters.
    \cite{Arteconi2016} perform an analysis to evaluate the benefits of DR programs in terms of electricity consumption and operational costs, both from the end user and the system perspective. The demand side technologies considered are electric heating systems, i.e. heat pumps and electric resistance heaters, coupled with thermal energy storage, i.e. the thermal mass of the building envelope and the domestic hot water tank.
    \cite{Chassin2016} develop a logistic demand curve for short term electricity consumption derived from the first principles of controllable thermostatic electric loads which corresponds to the random utility model commonly used in the economics of consumer choice. 
    \cite{Wang2013} use a bottom-up modeling framework for responsive spatially-distributed populations of heat pumps and other flexible loads into a security constrained economic dispatch formulation. Regional pockets of responsive loads are aggregated into models that describe population dynamics as an equivalent virtual power plant as a new source of spinning reserves. 
 
    \cite{Dhulst2015} present a flexibility estimation of residential loads including domestic hot water buffers based on measurements from a demonstration project in Belgium. They compute the maximal amount of time a certain increase or decrease of power can be realized within the comfort requirements of the user on appliance level. However, this flexibility potential varies during the day, and the potential for increasing or decreasing the power consumption is in general not equal. For wet appliances, an average maximum potential is presented that depends on the time of the day.
    \cite{Yin2016} present a demand response estimation framework using two-state models for thermostatically controlled loads in commercial and multi-dwelling residential units. Regression models are fit to a large dataset to predict the flexibility potential on unit level based on key inputs, including hour of day, set point change and outside air temperature.    

    The most commonly used thermostat control variable in heating, ventilating, and air conditioning (HVAC) systems is the setpoint of indoor air temperature. However, people's thermal comfort responds to operative temperature more directly than air temperature. 
    Based on three heating and cooling systems in three different geographical locations, \cite{Wang2019a} investigate how the adoption of operative temperature based control would affect the energy use. The authors show that the impact of the control variable strongly depends on the location and heating system type.
    \cite{Wang2019} provide an overview of research for HVAC systems in non-residential buildings to provide frequency regulation. 
    \cite{Tabares-Velasco2019} develop an optimization model to minimize electricity cost and user discomfort. The framework uses a model predictive control formulation capable of reducing cooling electricity costs by up to 30\%.
    \cite{Zhao2015} outline a high-level supervisory control strategy that directs interdependent HVAC systems of large commercial buildings for frequency reserves.
    \cite{Lakshmanan2016} study the provision of secondary frequency from TCLs and quantify the computation resource constraints for the control of a large TCL population. They then conduct an experimental investigation with domestic fridges representing in an islanded power system to evaluate the control. The experimental results show that TCLs are fast responsive loads for DR activation, with the average control signal response time of 24 seconds and an equivalent ramping rate of 63\% per minute, which could also comply with the requirements for primary frequency control. 
    
    Primary frequency control\footnote{Examples for primary frequency control products are frequency containment reserve (FCR) in continental Europe, firm frequency response in Great Britain, and 10 min spinning reserve in North America.} imposes more challenging requirements on the response time, and not all heat pumps may be available at the time of activation. This is due to deadband restrictions on the switching of the compressor.
    However, \cite{Muhssin2018} demonstrate that the aggregation of heat pumps and fridges offered large power capacity and, therefore, an instantaneous frequency response service is achievable. Specifically, the firm frequency response as used in Great Britain is investigated.
    \cite{Muller2019} illustrate that load reductions of 40–65\% of the total load can be achieved by throttling heat pumps, and that these load reductions can be delivered precisely with a median absolute percentage error of below 7\%, based on a demand response demonstration involving a population of more than 300 residential buildings with heat pumps. 
    
    \begin{table*}[t]
    \centering
    \caption{Modeling approaches in the related literature}
    \begin{tabular}{l cc ccc}
    \hline
    Reference                   & Service & Scale & Climate Zone \\ \hline
    \cite{Campillo2012}         & energy use          & several countries & no \\
    \cite{Mahdavi2016}          & energy use          & 10,000 ACs        & no \\
    \cite{Yin2016}              & energy use          & < 1,000 TCLs  & \textbf{considered} \\
    \cite{Wang2019a}            & energy use          & 1 TCL             & \textbf{considered} \\
    \cite{Wang2019}             & ancillary services  & comparison        & no \\
    \hline
    \cite{Tabares-Velasco2019}  & EA                  & 1 AC              & no \\
    \cite{Winstead2020}         & EA                  & 120 TCLs          & no \\
    \cite{Chassin2016}          & EA                  & < 1,000 TCLs      & no \\
    \cite{Dhulst2015}           & EA                  & 15 domestic hot water buffers & no \\
    \cite{Arteconi2016}         & EA                  & domestic hot water & no \\
    \cite{Mathieu2015}          & EA                  & 1,000 ACs         & no \\
    \cite{Muller2019}           & load reduction      & 300 HPs           & no \\
	\cite{Wang2013}             & spinning reserve    & 3,800 HPs         & no, but reserve zones \\
    \cite{Muhssin2018}          & firm frequency response & > 1,5$\cdot 10^6$ HPs & no \\
    \cite{Zhao2015}             & frequency regulation& 1 AC              & no \\
    \cite{Lakshmanan2016}       & secondary frequency control & refrigerators & no \\
    \cite{Callaway2009}         & regulation reserve  & 10,000 TCLs       & no \\
	\cite{Herre2020TCL}         & \textbf{EA \& FCR}& 1,000 ACs         & no \\
    \cite{Mathieu2014dk}        & EA                  & \textbf{national} (DK)  & no \\
	\cite{Nyholm2016}           & EA                  & \textbf{national} (SE)  & no \\
    \cite{MathieuCalifornia2015}& ancillary services  & 1,000 TCLs        & \textbf{considered} \\
    \hline
    \textbf{This paper}         & \textbf{EA \& FCR}& \textbf{national} (SE)  & \textbf{considered} \\
    \hline
    \multicolumn{4}{l}{\small AC: air conditioner, EA: energy arbitrage, HP: heat pump, FCR: frequency containment reserve.}
    \end{tabular} 
    \label{tab:Literature}
    \end{table*}
    
    The relevant literature on the optimization, control, and flexibility estimation of TCL is classified in \cref{tab:Literature} with respect to the type of power system service, the number of appliances, and the type of climate zone model.
	Previous studies mostly focus on the potential of buildings to shift energy demand to periods of low electricity prices. 
	The potential from electric space heating of Swedish SFDs is investigated with a detailed model in \cite{Nyholm2016}. The objective is to minimize the energy cost of each building by arbitraging energy prices and while maintaining an acceptable indoor thermal comfort. 
	The potential demand response (DR) capacity from controllable loads including TCLs in Denmark has been investigated in \cite{Mathieu2014dk}. The maximum coincident power (MCP) and daily shiftable energy (DSE) are used as key measures of DR potential that vary with time and ambient temperature. These measures are useful tools to assess the shift of energy, but less useful for to quantify the flexibility for ancillary services.
	TCLs have been proposed for energy arbitrage \citep{Nyholm2016} and ancillary services \citep{MathieuCalifornia2015}. However, \cite{Alam2020} show that the joint provision of multiple services yields increased economic benefits. To that end, a population of cooling TCLs providing multiple services 
	is investigated in \cite{Herre2020TCL}. 
    
	The cost-optimal power consumption schedule of a TCL population depends on energy prices. The cost-optimal power capacity offered to the reserve market depends on reserve prices and the time for which actions need to be sustained. Both depend on ambient temperature, which affects TCL operation \citep{Mahdavi2016}, availability (e.g., heaters are only available for control if it is sufficiently cold outside and they are powered on), and market timing. Market timing parameters include \textit{lead time}, i.e., the time between gate closure and operation, and \textit{contract period}, i.e., the period for which a service is committed. Furthermore, the power consumption schedule impacts the feasible reserve capacity and so the DA energy and reserve self-scheduling problems should be solved together.
	
	In this paper, we jointly optimize energy and reserve bids in order to explore the impact of market timing parameters on the business case of TCL aggregators, specifically on the profit and flexibility in terms of the reserve capacity offered to the system operator.
	We first formulate a rolling horizon optimization (RHO) problem that maximizes aggregator profit subject to uncertainty in real-time prices and in the activation signal for reserves. Additionally, uncertainty in the availability of TCLs and consequently in their power and energy bounds is represented by chance constraints.
	Since we are interested in developing a qualitative understanding of the relationship between profit/flexibility and market timing, we use a simplified thermal energy storage model of the TCL aggregation dynamics proposed in \cite{Mathieu2015}.
	
	The main contribution of this paper is to reveal the order of magnitude of the technical DR potential from the heating sector that are electrically heated in Sweden. We provide insight to the magnitude of reserves available from TCLs in (1) today's Swedish market setup, and in (2) an envisaged market with delayed gate closure and shorter contract periods. Specifically, the contributions are as follows:
	\begin{itemize}
	    \item We build an inventory of the entire Swedish single- and two-family dwelling (SFD) stock, heating types, insulation properties, and construction year. This extensive survey constitutes the foundation for performing the analysis in this paper.
	    \item We adapt the methodology of \cite{Mathieu2015} for computing the resource potential of TCLs based on the \emph{available} and \emph{capable} TCLs at a given ambient temperature.
	    \item Our methodology for the joint optimization of multiple services builds on that proposed by \cite{Herre2018PSCC} and \cite{Herre2020TCL}, which conducted a similar study for New England. However, here, we include both day-ahead and real-time energy which leads to a different problem formulation and accordingly different insights from the case studies. Furthermore, we model the entire Swedish SFD stock as thermal energy storages in a multi-area formulation.
	\end{itemize}
	
	
	Modeling differences with respect to \cite{Mathieu2015} and \cite{Herre2020TCL} will be discussed in \cref{sec:Method}, which details our methods. 
	\cref{sec:Data} outlines the available Swedish data and \cref{sec:Case} presents the results of our numerical investigation. We conclude in \cref{sec:Conclusion} with policy implications.

\section{Method} \label{sec:Method}

We assume that a TCL aggregator is capable of dynamically controlling a TCL population and has perfect foresight of ambient temperatures.
The indoor temperature measurements $\theta_{b,t}^{\text{in}}$ at time $t$ of building $b$ are automatically collected by the aggregator.
The aggregator is capable of sending a control signal in the form of \emph{on}/\emph{off} control.
We assume that TCLs have a constant efficiency that relates the thermal and electric power capacity with $P^{th}_{a,b,h}=P^{el}_{a,b,h} \eta_{h}$. The $b^\text{th}$ building can physically be described by the thermal resistance $R_b$ and capacitance $C_b$ in building $b$. 
We introduce first an individual model that represents the physical behavior of a TCL, and then an aggregate model for centralized control that has been proposed to efficiently control a large population \citep{Mathieu2013}.

\subsection{Individual Dwelling Model} \label{sec:Ind}
The evolution of the indoor temperature $\theta_{t,b}^{\text{in}}$ at time $t=t_0,\dots,t_{N}$ can be derived from a differential equation and discretized over time step $\Delta t$ using the model developed by \cite{Ihara1981} and extended to heterogeneous populations by \cite{Ucak1998},
\begin{equation}
    \theta^{\text{in}}_{b,t+1} = \kappa_b \cdot \theta_{b,t}^{\text{in}} + \left( 1 - \kappa_b \right) \cdot \left( \theta_{b,t} + m_{b,t} R_b P_b^{el} \eta_h \right) + \gamma
    \label{eq:PhyM}
\end{equation}
where $\kappa_b=\text{exp}(-\frac{\Delta t}{R_b C_b})$ and the noise term $\gamma$ is commonly neglected \cite{Callaway2009}. The indoor temperature set point $\theta^\text{set}_b$ and bounds $\overline{\theta}_b$ and $\underline{\theta}_b$ are specified by the user and can be communicated to the aggregator. The uncontrolled operation is defined by the \emph{on}/\emph{off} state $m_{b,t}$ where
\begin{equation}
    m_{b,t} = \begin{cases} 
        1, & \text{if}~ \theta_{b,t}^\text{in} < \underline{\theta}_b,\\ 
        0, & \text{if}~ \theta_{b,t}^\text{in} > \overline{\theta}_b,\\
        m_{b,t-1}, & \text{otherwise.} 
        \end{cases}
    \label{eq:mOnOff}
\end{equation}

\subsection{Thermal Energy Storage Model of Dwelling Stock} \label{sec:TES}
For a heating TCL, the availability is defined analogously to \cite{Mathieu2015} in \cref{eq:avail}. A heating TCL of type $h$ in building $b$ is available if it is sufficiently cold, and if the TCL is able to heat the space to within the deadband.
\begin{subequations}
  \begingroup
  \allowdisplaybreaks
  \begin{align}
    & \hat{v}_{a,b,h,t} = \begin{cases} 
    1 & \text{if}~ \theta_{a,t} < \underline{\theta}_{b,h} \\
    0 & \text{otherwise}
    \end{cases} \label{eq:avail1}\\
    & \tilde{v}_{a,b,h,t} = \begin{cases} 
    1 & \text{if}~ \theta_{a,t} + R_{b} \cdot P^{el}_{a,b,h} \cdot \eta_{h} > \overline{\theta}_{b,h}  \\
    0 & \text{otherwise}
    \end{cases} \label{eq:avail2}
  \end{align}
  \endgroup
  \label{eq:avail}
\end{subequations}
The \emph{ON} \cref{eq:tOn} and \emph{OFF} \cref{eq:tOff} time of a heating TCL are obtained from equation \cref{eq:PhyM}, and its duty cycle is given by equation \cref{eq:dOnOff}. 
\begin{subequations}
  \begingroup
  \allowdisplaybreaks
  \begin{align}
  T^{ON}_{a,b,h,t}   &= C_b R_b \cdot ln \left( \frac{ \underline{\theta}_{h,b}-\theta_t - R_b P^{el}_{a,b,h} \eta_h }{ \overline{\theta}_{b,h}-\theta_t - R_b P^{el}_{a,b,h} \eta_h} \right) \label{eq:tOn} \\
  T^{OFF}_{a,b,h,t}  &= C_b R_b \cdot ln \left( \frac{\overline{\theta}_{b,h}-\theta_t}{\underline{\theta}_{b,h}-\theta_t } \right) \label{eq:tOff} \\
  D_{a,b,h,t}     &= \frac{T^{ON}_{a,b,h,t}}{T^{ON}_{a,b,h,t}+T^{OFF}_{a,b,h,t}}  \label{eq:dOnOff}
  \end{align}
  \endgroup
  \label{eq:OnOff}
\end{subequations}
The aggregate parameters of a heating TCL population at ambient temperature $\theta_{a,t}$ can be computed offline for all $a$ and $t$ as below, which is extended from \cite{Mathieu2015}.
\begin{subequations}
  \begingroup
  \allowdisplaybreaks
  \begin{align}
    P^B_{a,t}       &= \sum_{h\in\mathcal{H}} \sum_{b\in\mathcal{B}} D_{a,b,h} \cdot P^{el}_{a,b,h}   \\
    \overline{P}_{a,t}  &= \sum_{h\in\mathcal{H}} \sum_{b\in\mathcal{B}} \hat{v}_{a,b,h,t} \cdot P^{el}_{a,b,h}    \\ 
    \underline{P}_{a,t} &= \sum_{h\in\mathcal{H}} \sum_{b\in\mathcal{B}} \tilde{v}_{a,b,h,t} \cdot P^{el}_{a,b,h} \\
    \overline{S}_{a,t}  &= \sum_{h\in\mathcal{H}} \sum_{b\in\mathcal{B}} P^{el}_{a,b,h} \cdot T^{ON}_{a,b,h,t} \cdot (1-D_{a,b,h,t}) \\
    \underline{S}_{a,t} &= 0
  \end{align}
  \endgroup
  \label{eq:paramsPopL}
\end{subequations}
The difference to the aggregate parameters in \cite{Mathieu2015} is that, here, the upper and lower bounds describe the \emph{entire} TCL population at all times. The bounds in \cite{Mathieu2015} aggregate only the \emph{available and capable} TCLs in the population, i.e., those instances where $\hat{v}_{a,b,h,t}=1$ and $\tilde{v}_{a,b,h,t}=1$.

\subsection{Mathematical Formulation of Day-Ahead Bidding} \label{sec:Formulation}

The objective of the aggregator is to maximize its profit $\mathrm{\Pi}_{\omega}$ which is comprised of revenue from reserve capacity $p_{a,t}^{R}$ and cost from power consumption $p_{a,t,\omega}^{RT}$, where $\lambda_{t}^{R},\lambda_{a,t}^{DA}$ and $\lambda_{a,t,\omega}^{RT}$ are the reserve, DA, and RT energy prices, respectively. 
	
The aggregator faces uncertainty from both prices and TCL availability. We propose a method to manage these uncertainties simultaneously. Specifically, we formulate a two-stage chance-constrained optimization problem in which the chance constraints \cref{eq:StoP1T,eq:StoP2T,eq:StoE1T,eq:StoE2T,eq:Stoc1T,eq:Stoc2T} include second-stage variables that depend on price realizations in each scenario $\omega$. There are different ways to implement the chance constraints. Since we assume the aggregator is primarily concerned about service quality and feasibility, we use a conservative approach in \cref{eq:StoAllT} which requires that in each price scenario and in each market interval there is at least a $1-\epsilon_i$ chance that the service quality is met. 
The  profit maximization problem with decision variables $\mathbf{x} = \lbrace c_{a,\omega},\, c_a^{R},\, c_a^{DA},\, c_{a,\omega}^{RT},\, c_{a,\omega}^{I},\, c_{a,\omega}^{P},\, e_{a,t,\omega}^{I},\\ e^{DA}_{a,t},\, p_{a,t}^{R},\, e_{a,t,\omega}^{R},\, p_{a,t,\omega}^{RT},\, e^{RT}_{a,t,\omega},\, s_{a,t,\omega},\, d_{a,\omega},\, z_{a,\omega},\, y_a,\, \text{CVaR}_{a}^{\alpha} \rbrace$ is set out in \cref{eq:StoAllT}
\begin{subequations}
  \begingroup
  \allowdisplaybreaks
  \begin{align}
  	&  \hspace{-.2cm} \underset{\mathbf{x}}{\text{max}}. \; 
	\left( 1-\beta \right) \mathbb{E} \left[ \sum_{a\in\mathcal{A}} \mathrm{c}_{a,\omega} \right] + \beta \; \sum_{a\in\mathcal{A}} \text{CVaR}_{a,\alpha}  
	\label{eq:StoObjT}\\
	&  \hspace{-.2cm}\text{s.t.}\;    \mathrm{c}_{a,\omega} = c_a^{R} - c_a^{DA} - \left( c_{a,\omega}^{RT} + c_{a,\omega}^{I}+ c_{a,\omega}^{P} \right)   \;\forall a,\omega
    \label{eq:StoProfT}\\
    &    c_a^{R} = \sum_{t=\tau}^{\tau+T_c} p_{a,t}^{R} \cdot \lambda_t^{R}  \;\forall a
    \label{eq:revR}\\
    &    c_a^{DA} = \sum_{t=\tau}^{\tau+T_c} e_{a,t}^{DA} \cdot \lambda_{a,t}^{DA}  \;\forall a
    \label{eq:costDA}\\
    &    c_{a,\omega}^{RT} = \sum_{t=\tau}^{\tau+T_c} \left( e^{RT}_{a,t,\omega} - e^{DA}_{a,t} \right) \cdot \lambda_{a,t,\omega}^{RT}   \;\forall a,\omega
    \label{eq:costRT}\\
    &    c_{a,\omega}^{I} = \sum_{t=\tau}^{\tau+T_c} e_{a,t,\omega}^{I}\cdot \lambda^{I}   \;\forall a,\omega
    \label{eq:costItot}\\
    &    e_{a,t,\omega}^{I} \geq ~~~~ e^{RT}_{a,t,\omega} - e^{DA}_{a,t} \;\forall a,t,\omega
    \label{eq:costIa}\\
    &    e_{a,t,\omega}^{I} \geq - \left( e^{RT}_{a,t,\omega} - e^{DA}_{a,t} \right) \;\forall a,t,\omega
    \label{eq:costIb}\\
    &    c_{\omega}^{P} = \sum_{a\in\mathcal{A}} d_{a,\omega}\cdot \lambda^{P}   \;\forall a,\omega
    \label{eq:costP} \\	
    &    0 \le e_{a,t}^{DA} \le \overline{P}_{a}^\text{inst} \cdot \Delta t   \;\forall a,t
    \label{eq:eDA}\\    
    &    0 \le p_{a,t,\omega}^{RT} \le \overline{P}_{a}^\text{inst} \cdot \Delta t   \;\forall a,t,\omega
    \label{eq:pRT}\\    
    &    e_{a,t,\omega}^{RT} = p_{a,t,\omega}^{RT} \cdot \Delta t   \;\forall a,t,\omega
    \label{eq:peE}\\    
    &    p_{a,t}^{R} \ge 0  \;\forall a,t
    \label{eq:pR}\\    
    &    e_{a,t,\omega}^{R} = p_{a,t}^{R} \cdot \tfrac{f_{t,\omega} - \hat{f}}{0.1} \cdot \Delta t   \;\forall a,t,\omega
    \label{eq:peR}\\
    &    \mathbb{P} \{ p_{a,t,\omega}^{RT} + p_{a,t}^{R} \le \overline{\mathcal{P}}_{a,t} \} \ge 1-\epsilon_1   \;\forall a,t,\omega
	\label{eq:StoP1T} \\
	&    \mathbb{P} \{ p_{a,t,\omega}^{RT} - p_{a,t}^{R} \ge \underline{\mathcal{P}}_{a,t} \} \ge 1-\epsilon_2   \;\forall a,t,\omega
	\label{eq:StoP2T} \\
	&    \mathbb{P} \{ s_{a,t,\omega} + \tfrac{\Delta t}{2} p_{a,t}^{R} \le \overline{\mathcal{S}}_{a,t} \} \ge 1-\epsilon_3   \;\forall a,t,\omega
	\label{eq:StoE1T} \\
	&    \mathbb{P} \{ s_{a,t,\omega} - \tfrac{\Delta t}{2} p_{a,t}^{R} \ge \underline{\mathcal{S}}_{a,t} \} \ge 1-\epsilon_4    \;\forall a,t,\omega
	\label{eq:StoE2T}\\
	&    \mathbb{P} \{ s_{a,\tau_h,\omega} - d_{a,\omega} \le \mathcal{G}_{a,\tau_h} \} \ge 1-\epsilon_5   \;\forall a,\omega 
	\label{eq:Stoc1T}\\
	&    \mathbb{P} \{ s_{a,\tau_h,\omega} + d_{a,\omega} \ge \mathcal{G}_{a,\tau_h} \} \ge 1-\epsilon_6   \;\forall a,\omega 
	\label{eq:Stoc2T}\\
	&    d_{a,\omega} \ge 0   \;\forall a,\omega
	\label{eq:Stoc3T}\\
	&    s_{a,t,\omega} = s_{a,t-1,\omega} + e_{a,t,\omega}^{R} + (p_{a,t,\omega}^{RT} - P^{B}_{a,t}) \Delta t   \;\forall a,t,\omega
	\label{eq:StoSOCT} \\
	&    s_{a,\tau-1,\omega} = {G}_{a,\tau-1}   \;\forall a,\omega 
	\label{eq:StoInitT} \\
	&    y_a - \tfrac{1}{(1-\alpha)} \sum_{\omega\in\Omega} \pi_\omega v_{a,\omega} = \text{CVaR}_{a}^{\alpha}   \;\forall a,\omega
	\label{eq:StocCVaRT} \\
	&    y_a - \mathrm{c}_{a,\omega} \leq v_{a,\omega}   \;\forall a,\omega
	\label{eq:StocEtaT} \\
	&    v_{a,\omega} \geq  0  \;\forall a,\omega 
	\label{eq:StocIotaT}
  \end{align}
  \endgroup
  \label{eq:StoAllT}
\end{subequations}
	where $\omega$ denotes a scenario in set $\Omega$.
	Constraints \cref{eq:StoP1T,eq:StoP2T,eq:StoE1T,eq:StoE2T} model the TCL aggregation power and energy limits assuming worst-case reserve activation, which occurs when the reserve signal is equal to $ \pm p_{a,t}^{R}$ during the first $\frac{\Delta t}{2}$ and equal to $\mp p_{a,t}^{R}$ during the second $\frac{\Delta t}{2}$. 
	Constraints \cref{eq:Stoc1T,eq:Stoc2T,eq:Stoc3T} 
	together with the last term in the objective function penalize the deviation $d_\omega$ of the final SOC within the optimization horizon $\tau_h$ from 50\%, where $\mathcal{G}_{a,t} = \tfrac{1}{2} \left( \overline{\mathcal{S}}_{a,t} + \underline{\mathcal{S}}_{a,t} \right)$ and $\lambda^P$ is the deviation penalty. Constraint \cref{eq:StoSOCT} models the evolution of the SOC and \cref{eq:StoInitT} sets the initial SOC to 50\%. 
	
	We use a RHO approach, where the optimization horizon includes market intervals $t=\tau, \dots,\tau_h$ with $\tau_h=\tau+T_c+T_p$. The contract time $T_c$ of the day-ahead energy market and the reserve market is 24\,h, and the extended prediction horizon $T_p$ can be selected by the aggregator.
	
	We use the methods detailed in \cite{Herre2018PSCC} to mitigate the propagation of baseline forecast error and obtain cumulative distribution functions (CDFs) for the random TCL parameters. This allows us to reformulate the chance constraints as deterministic linear constraints as suggested in \cite{Kataoka1963}. The resulting problem is a linear program. Compared to the problem in \cite{Herre2020TCL}, this formulation includes (i) a multi-area formulation, (ii) the fact that the activation signal may not have a zero mean energy character (equation \cref{eq:peR}), and (iii) both the day-ahead (DA) and the real-time (RT) market. 
	This formulation therefore allows to capture not only energy arbitrage between different times, but also energy arbitrage between price differences in DA and RT energy markets.

\section{Data Sources and Processing}\label{sec:Data}
\begin{figure}[t]
\centering
\includegraphics[width=0.49\textwidth]{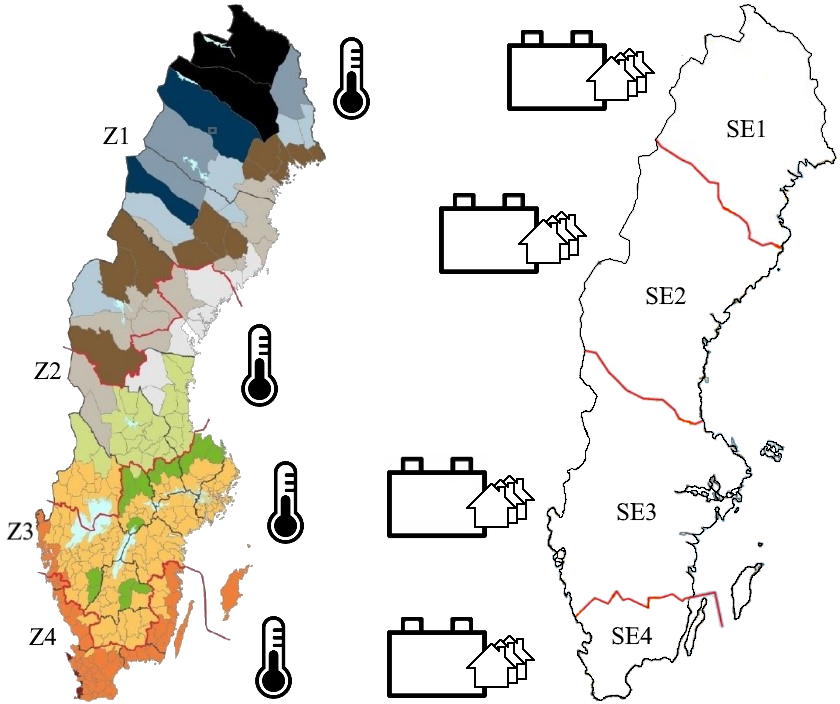}
\caption{Climate zones $z$ (left) and price areas $a$ (right) in Sweden. In the left figure, red lines represent borders of climate zones $z$ and grey lines represent borders of counties $k$.}
\label{fig:ClimateZones}. 
\end{figure}

This section summarizes the input data consisting of \; Swedish (a) climate and temperature data, (b) building stock data, and (c) power system data. \cref{fig:ClimateZones} illustrates the high-level approach of this study. The left side shows the four climate zones, and the right side shows the four price areas in Sweden. Here, we assume that the mean temperature in zone $z$ represents the ambient conditions in price area $a$ sufficiently well. Furthermore, building stock data from 21 counties is aggregated by price area and used to quantify the equivalent thermal energy storage parameters of price area $a$. 
The data acquisition and processing is detailed in the remainder of this section.

\subsection{Temperature Data}\label{sec:TempData}
Outdoor temperatures are obtained as long-term hourly mean temperature measurements in all counties. We obtained 20 years of historical hourly temperature measurements from \cite{Meteo2020}. The most important design parameters of the temperature data are summarized in \cref{tab:ClimateZones} per county.

\begin{figure*}[t]
\centering
\includegraphics[width=\textwidth,trim={0.2cm 0.7cm 0.2cm 0.2cm},clip]{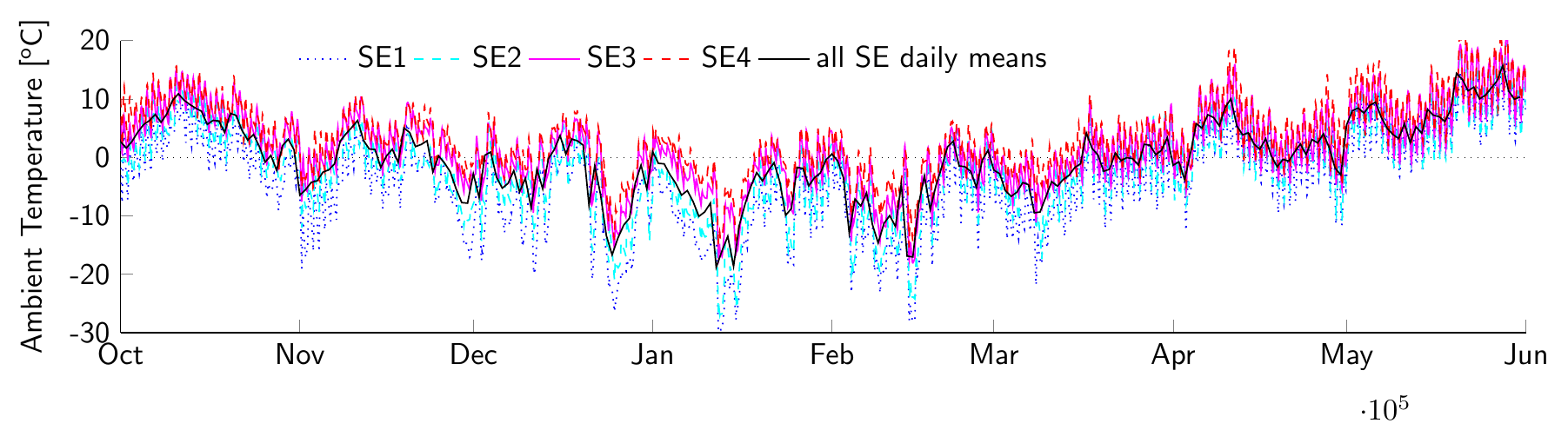}
\caption{Ambient temperatures given as long (20 year) term hourly means in each price area. The average temperature over all Sweden is given as daily mean values in black.}
\label{fig:DataTemp}
\end{figure*} 
Since we are interested in the flexibility potential in each price area, we use the weighted mean hourly temperature per price area. The weight is the share of SFDs installed in the respective county. \cref{fig:DataTemp} shows the long-term hourly mean temperatures in each price area that are used in \cref{sec:Case}. It can be seen that the average temperatures can differ more than 20$^o$C between the north (SE1) and the south (SE4).

\subsection{Climatic Zones and Heating Design}
The Swedish National Board of Housing, Building and Planning, Boverket, provides general recommendations and mandates for the energy use of Swedish buildings in \cite{Boverket2018}. According to the latest building regulation release, Sweden is divided into four climate zones. Zones $Z1$ to $Z4$ in \cref{fig:ClimateZones} divide Sweden into cold northern and mild southern zones. According to \cite{Boverket2018}, the design of electrically heated buildings needs to comply with the allowable specific energy use for heating and ventilation (HV) in each climate zone, i.e., $E_z^{HV}=[95, 75, 55, 50] \tfrac{\text{kWh}}{\text{m}^2\,\text{year}}$ for zones $z_1$ to $z_4$. 

The heating requirements of each specific county $k$ can be quantified by heating degree days ($\theta_z^\text{DD}$) as listed in \cref{tab:ClimateZones}. 
Heating degree days quantify the number of degrees that the daily average temperature is below the setpoint for heating, multiplied by $24\,\frac{\text{h}}{\text{day}}$ and summed over a year, in the respective county. The maximum electrical power is drawn when electric heating, DHW preparations, and ventilation occur simultaneously, at the peak heating demand. The peak demand can be calculated at the design ambient temperature $\theta_z^\text{des}$ of the respective county. The yearly median ambient temperature is given by $\theta_z^\text{med}$.
Using the average dwelling surface $A^\text{SFD}=122\,\text{m}^2$ per SFD \citep{SCBdwellingstock2016}, the thermal energy required for heating and ventilation purposes $Q_{h,z}^\text{HV}$ can be calculated as: 
\begin{equation}
  	Q_{h,k}^\text{HV} = E_k^\text{HV} \cdot \eta_h \cdot \frac{ \theta^\text{set}-\theta_k^\text{des} }{ \theta_k^\text{DD} } \cdot A^\text{SFD} \; \left[\tfrac{\text{kWh}}{\text{year}}\right] \; \forall h, k
\label{eq:calcEnergy}
\end{equation}
where $h\in \mathcal{H}$ is the heating type and $k$ is the county. For simplicity, and without loss of generality, the thermal efficiency $\eta_h \approx \eta_{h,k}$ of heating system $h$ is assumed constant throughout all climate zones and counties in the case study.

\begin{table}[t]
  \centering
  \caption{Climatic zones and obtained power ratings by county}
    \setlength{\tabcolsep}{3pt}
    \begin{tabular}{ll rrrrrrrr}
    \hline
    $k$ & County              &$\theta_k^\text{des}$&Area	& $\theta_k^\text{med}$& $\theta_k^\text{DD}$&P$_\text{gs}^\text{el}$& P$_\text{aa}^\text{el}$ & [\%]\\
    \hline 
    1 & Norrbotten          &   -30.0&  1& -1.2&	144,000&	6&	12& 2 \\
    2 & V{\"a}sterbotten    &   -26.3&	1&  3.4&	120,500&	6&	10& 4 \\
    \hline
    3 & J{\"a}mtland        &	-23.5&	2&	2.7&	128,000&	5&	10& 2 \\
    4 & V{\"a}sternorrland  &	-24.7&	2&	2.6&	129,000&    5&	10& 3 \\
    5 & V{\"a}rmland        &	-21.0&	2&  5.9&    101,000&	6&	10& 4 \\
    6 & Dalarna             &	-20.8&	2&	5.2&	106,000&	6&	10& 4 \\
    7 & G{\"a}vleborg       &	-18.2&	2&	5.7&	105,000&	5&	10& 4 \\
    \hline
    8 & {\"O}rebro          &	-19.1&	3&	5.9&	101,000&	4&	10& 4 \\
    9 & V{\"a}stmanland     &	-18.7&	3&	5.6&	103,000&	4&	 6& 2 \\
    10& Uppsala l{\"a}n	    &   -17.9&	3&	  6&  	 99,500&	4&	 6& 4 \\
    11& S{\"o}dermanland    &	-17.6&	3&	  6&	 99,500&	4&	 6& 3 \\
    12& {\"O}sterg{\"o}tland& 	-16.6&	3&	6.1&	 98,800&	4&	 6& 4 \\
    13& J{\"o}nk{\"o}ping   &   -16.1&	3&	6.5&	 95,500&	4&	 6& 4 \\
    14& Stockholm 	        &   -15.9&	3&  6.6&     95,000&	4&	 6& 14 \\
    15& V{\"a}stra G{\"o}taland&-13.6&	3&	7.9&	 83,500&	4&	 6& 17 \\
    16& Gotland             &	 -9.4&	3&	  8&	 82,700&	4&	 6& 1 \\
    \hline
    17& Kronoberg 	        &   -15.0&	4&  6.9&	 92,000&	4&	 6& 3 \\
    18& Kalmar              &	-14.5&	4&	  7&	 91,400&	4&	 6& 3 \\
    19& Halland             &	-14.3&	4&	7.7&	 85,000&	4&	 6& 4 \\
    20& Sk{\aa}ne	        &   -11.0&	4&	  8&	 82,700&	4&	 5& 13 \\
    21& Blekinge	        &   -10.9&	4&	  8&	 82,700&	4&	 5& 3 \\
    \hline 
    \end{tabular} 
  \label{tab:ClimateZones}%
\end{table}%
\begin{table}[t]
  \centering
  \caption{Rating and thermal efficiency of electric heating methods}
    \setlength{\tabcolsep}{4pt}
    \begin{tabular}{l rrrr r}
    \hline
                                    & \multicolumn{4}{c}{$P^{el}_{a,h}$ [kW]} & Efficiency\\
    Heating type $h$                & $a_1$& $a_2$& $a_3$& $a_4$   & $\eta_h$\\
    \hline 
    $h_1$ ground source heat pump ($H^{gs}$)  &    6 &    5 &    4 &    3  & 3.0\\
    $h_2$ air-to-air heat pump ($H^{aa}$)     &   12 &   10 &    6 &    5  & 2.5\\
    $h_3$ air-to-water heat pump ($H^{aw}$)   &   12 &   10 &    6 &    5  & 2.3\\
    $h_4$ exhaust air heat pump ($H^{ex}$)    &   12 &   10 &    6 &    5  & 2.4\\
    $h_5$ direct electric heating ($H^{e}$)   &   18 &   15 &   12 &   10  & 0.98\\
    $h_6$ electric water boiler ($H^{w}$)     &   18 &   15 &   12 &   10  & 0.98\\
    \hline 
    \end{tabular} 
  \label{tab:hprating}%
\end{table}%
We follow the design rules of \cite{Boverket2018} and \cite{Tangix} to obtain the electric power rating of ground source ($H^{gs}$) and air-to-air ($H^{aa}$) heat pumps from $Q_{h,z}^\text{HV}$. The obtained electric power ratings are listed in \cref{tab:ClimateZones}. 
Air-to water ($H^{aw}$) and exhaust air heat pumps ($H^{ex}$) are assumed to have the same rating as $H^{aa}$. Furthermore, direct electric ($H^e$) heaters and electric water boilers ($H^w$) are designed with similar thermal power rating as ground source heat pumps. 
The electric power ratings per price area $a$, as well as the share of heating type are summarized by price area in \cref{tab:hprating}. 
Instead of the climate zone classification, we divide the country into counties $k$ that are then clustered into price areas $a$ according to \cref{tab:ClimateZones}.

\subsection{Swedish Building Stock}\label{sec:SweSFDdata}
The scope of this paper focuses on single- and two-family dwellings (SFD). Building insulation data of 14 different types is obtained from \cite{SEA2020} (and supporting background documentation is obtained on request). 
The thermal capacitance and resistance $C_b$ and $R_b$ of building $b$ can be obtained with the background data provided in \cite{SEA2020}. 
We assume slab foundations in half of the building stock and $0.75$ air changes per hour (ACH) to obtain mean values of $C_b$ and $R_b$. 
\begin{table}[t]
  \centering
  \caption{Percentage of heat pumps and electric heating per decade \citep{Nilsson2019}}
    \setlength{\tabcolsep}{5pt}
    \begin{tabular}{r rrrrrr}
    \hline
    Heating            &$H^{gs}$&$H^{aa}$&$H^{aw}$&$H^{ex}$&$H^{e}$&$H^{w}$ \\
    Index $h$          & 1 & 2 & 3 & 4 & 5 & 6 \\
    \hline
    {~~~~\;$<$1930}& 0.20  & 0.19  & 0.06   & 0.04   & 0.18   & 0.11 \\
    {1931-1940} & 0.20  & 0.19  & 0.06   & 0.04   & 0.18   & 0.11 \\
    {1941-1950} & 0.20  & 0.19  & 0.06   & 0.04   & 0.18   & 0.11 \\
    {1951-1960} & 0.21  & 0.20  & 0.07   & 0.05   & 0.10   & 0.14 \\
    {1961-1970} & 0.21  & 0.20  & 0.07   & 0.05   & 0.10   & 0.14 \\
    {1971-1980} & 0.18  & 0.17  & 0.06   & 0.04   & 0.13   & 0.16 \\
    {1981-1990} & 0.14  & 0.13  & 0.04   & 0.03   & 0.39   & 0.09 \\
    {1991-2000} & 0.10  & 0.10  & 0.03   & 0.02   & 0.16   & 0.34 \\
    {2001-2010} & 0.10  & 0.10  & 0.03   & 0.02   & 0.20   & 0.31 \\
    {2011-2019}     & 0.16  & 0.15  & 0.05   & 0.04   & 0.06   & 0.39 \\
    \hline
    \end{tabular} 
  \label{tab:Heatingshare}%
\end{table}%
The shares of heat pumps installed in SFDs and built in different time periods are obtained from \cite{SCB2020} and are listed in \cref{tab:Heatingshare}. 
Furthermore, the nominal efficiency $\eta_h$, or coefficient of performance (COP) of the heating systems is provided in \cref{tab:Heatingshare} for each heating type and assumed constant in all areas.

\begin{table}[t]
  \caption{Number of houses built per decade \citep{SCB2020}, assumed share of SFD building type per decade, and thermal properties for each house type \citep{SEA2020}}
    \resizebox{\textwidth}{!}{%
    \setlength{\tabcolsep}{4pt}
    \begin{tabular}{l | rrrr | rrrrr rrrrr rrrr}
    \hline
    Construction  & \multicolumn{4}{c}{Number of SFDs in area $a$} & \multicolumn{14}{c}{Share of SFD building type with index $b$}\\
    Year          & \multicolumn{4}{c}{from \cite{SCB2020}} & \multicolumn{14}{c}{from \cite{SEA2020}}\\
                  &   1 &   2 &   3 &   4 & 1 & 2 & 3 & 4 & 5 & 6 & 7 & 8 & 9 & 10 & 11 & 12 & 13 & 14\\
    \hline
    {~~~~\;$<$1930}&16,858&85,564 & 198,227 & 114,980 & 1    & -    & -    & -    & -    & -    & -    & -    & -    & -    & -    & -    & -    & - \\
    {1931-1940} &  9,561 & 28,180 &  69,992 &  33,283 & 0.50 & 0.50 & -    & -    & -    & -    & -    & -    & -    & -    & -    & -    & -    & - \\
    {1941-1950} & 10,108 & 25,086 &  70,838 &  31,930 & -    & 0.90 & 0.10 & -    & -    & -    & -    & -    & -    & -    & -    & -    & -    & - \\
    {1951-1960} & 15,939 & 33,306 &  79,012 &  36,302 & -    & -    & 0.95 & 0.05 & -    & -    & -    & -    & -    & -    & -    & -    & -    & - \\
    {1961-1970} & 17,890 & 38,142 & 157,704 &  76,132 & -    & -    & 0.80 & 0.12 & 0.08 & -    & -    & -    & -    & -    & -    & -    & -    & - \\
    {1971-1980} & 30,785 & 66,327 & 223,539 & 106,940 & -    & -    & -    & -    & -    & 0.22 & 0.56 & 0.22 & -    & -    & -    & -    & -    & - \\
    {1981-1990} & 14,115 & 30,957 & 114,704 &  55,159 & -    & -    & -    & -    & -    & -    & -    & -    & 0.72 & 0.28 & -    & -    & -    & - \\
    {1991-2000} &  5,493 &  9,301 &  57,749 &  26,736 & -    & -    & -    & -    & -    & -    & -    & -    & 0.15 & 0.85 & -    & -    & -    & - \\
    {2001-2010} &  4,392 &  9,144 &  67,319 &  33,691 & -    & -    & -    & -    & -    & -    & -    & -    & {0.10} & {0.15} & {0.18} & {0.18} & {0.20} & {0.20} \\
    {2011-2019} &  3,130 &  6,127 &  44,147 &  20,331 & -    & -    & -    & -    & -    & -    & -    & -    & -                     & -                     & {0.25} & {0.25} & {0.25} & {0.25} \\
    {no data}   &  2,005 &  2,651 &   2,648 &    8,845 & & & & & & & & & & & & & & \\
    \hline
    $C_b$ in $\tfrac{\text{kWh}}{\text{\text{K}}}$ & & & & & 7.89 & 11.43 & 11.77 & 42.65 & 7.45 & 9.19 & 15.57 & 7.30 & 7.60 & 4.53 & 15.40 & 10.80 & 7.52 & 11.37 \\
    $R_b$ in $\tfrac{\text{K}}{\text{kW}}$         & & & & & 4.66 & 5.79  & 7.91  &  3.89 & 8.56 & 8.77 &  6.06 & 6.32 & 5.68 & 9.64 &  7.02 &  7.24 & 6.98 &  5.89 \\ 
    \hline
    \end{tabular}}
  \label{tab:SFDSE}%
\end{table}%
We can obtain the number of SFDs built per time period and per county $k$ from \cite{SCB2016}. This geographical data can be clustered to obtain the number of SFDs built per price area $a$ using the clusters provided in \cref{tab:ClimateZones}. The resulting number of constructed SFDs per price area and decade is given in \cref{tab:SFDSE}, along with the estimated fraction of SFD types per decade. Note that this fraction is based on expert opinions since no statistical data are available.

\begin{figure}[t]
  \centering 
  \includegraphics[width=0.55\textwidth,trim={0.2cm 0cm 0cm 0cm},clip]{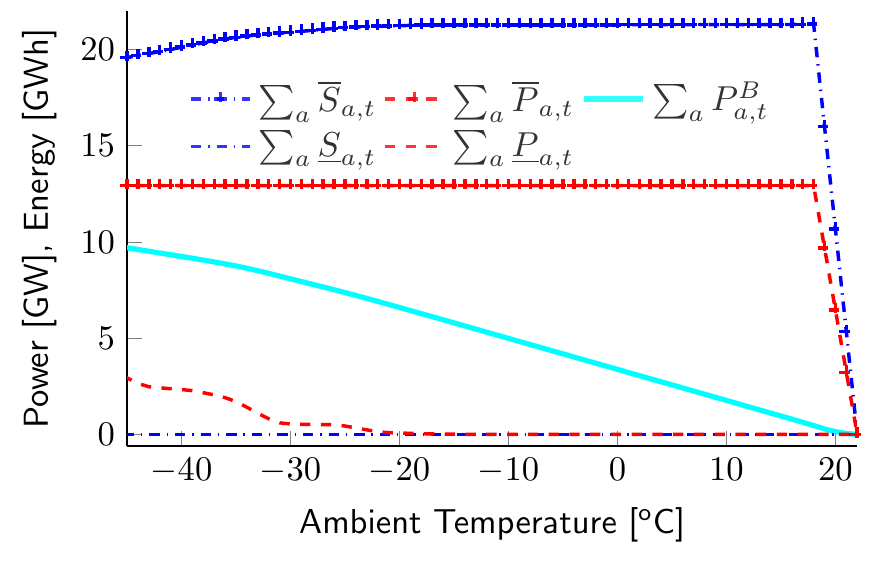}
  \caption{Thermal parameters of the population of electric heating and heat pumps over ambient temperature $\theta$, computed from equation \cref{eq:paramsPopL} from data in \cref{tab:hprating,tab:Heatingshare,tab:SFDSE}.}
  \label{fig:sensToutb}
\end{figure}

Finally, we can combine \cref{tab:hprating,tab:Heatingshare,tab:SFDSE} in order to obtain the number of SFDs per price area that are equipped with a given heating system, and whose building insulation can be quantified by $R_b$ and $C_b$.
With the methodology of \cref{sec:TES}, we can compute the aggregate parameters of heating TCLs in Sweden as shown in \cref{fig:sensToutb}

\subsection{Electricty Market Price \& Frequency Data}\label{sec:PriceData}
Hourly energy and real-time imbalance price data are obtained from \cite{NordPool} and reserve prices from \cite{Mimer}. A fixed conversion rate of $1\,\text{\textdollar}\!=\!10\,\text{SEK}$ is used to promote the understanding of the data and results. 
\begin{figure}[t]
\centering
\includegraphics[width=\textwidth,trim={0.2cm 0.7cm 0.2cm 0.2cm},clip]{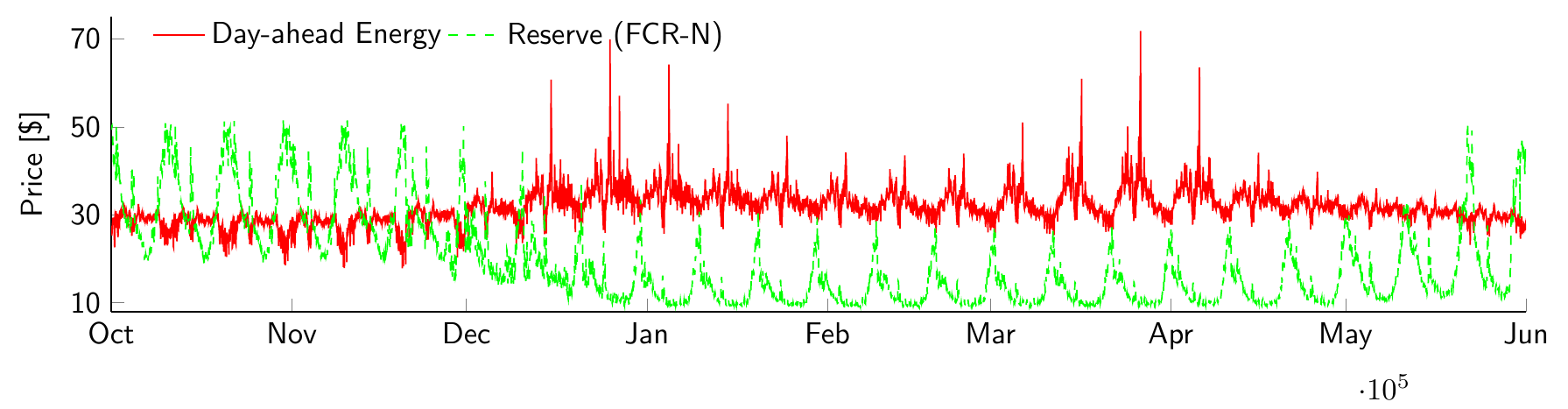}
\caption{Reserve prices [\textdollar/MW/0.1Hz] and day-ahead energy prices [\textdollar/MWh] as historical mean values (2013 to 2020).}
\label{fig:DataPrice}
\end{figure} 
\cref{fig:DataPrice} shows the average day-ahead energy and reserve prices. Note the difference in units; energy is traded in [\textdollar/MWh] per hour, while FCR-N reserves are traded in [\textdollar/MW/0.1Hz] per hour.
Almost half of the yearly electricity in Sweden is produced by hydropower stations. Their operation is highly affected by the water level in the reservoirs which depends on seasonal variable inflow. This is reflected in high variation of energy and reserve prices between the different periods of the year. For instance, the typical `spring flood' in May lowers electricity prices, and a dry summer or snowfall in winter (instead of rainfall) increases electricity prices. 

\begin{figure}[t]
\centering
\includegraphics[width=0.49\textwidth,trim={0.2cm 0.1cm 0.1cm 0.1cm},clip]{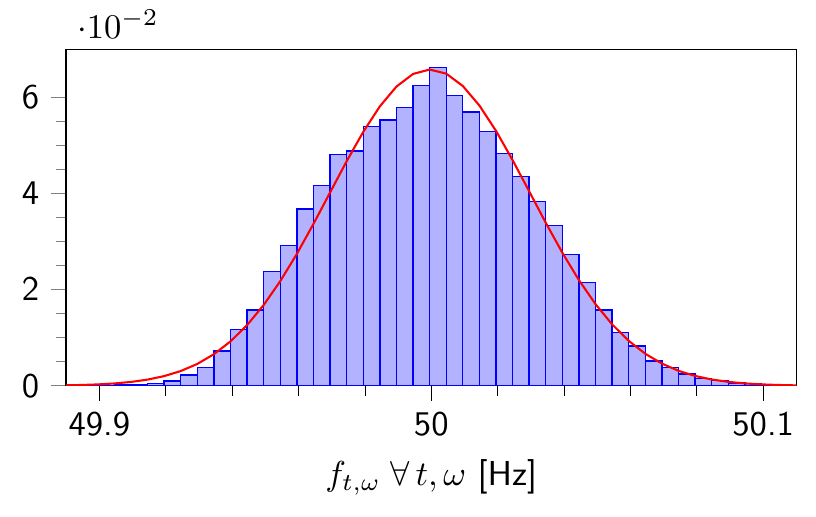}
\caption{Histogram of mean hourly frequencies from 2013 to 2020 and approximation by normal distribution.}
\label{fig:frequencyRT}
\end{figure}

Frequency data are available from Fingrid \cite{FingridFreq}. We use data from 2013 to 2020.
\cref{fig:frequencyRT} shows the probability density function of hourly mean frequencies. 

\subsection{Summary}\label{sec:DataSummary}
\begin{table}[t]
  \centering
  \caption{Summary of input data with corresponding sources}
    \resizebox{\textwidth}{!}{%
    \setlength{\tabcolsep}{2pt}
    \begin{tabular}{ll lll}
                               & &   & This paper (2021) & \cite{Nyholm2016} \\
    \hline
    \multicolumn{2}{c}{ \multirow{14}{*}{ \rotatebox{90}{ \textbf{Building stock}} } }
       & Number of SFDs with electric heating & 1.44 million (of 2.1 million in 2019) \citep{SCB2020} & 1.26 million (of 2.0 million in 2012) \\
     & & Number of building types        & 14 \citep{SEA2020} & 574 \citep{Boverket2011} \\
     & & Electricity demand for space heating & 19.6\,TWh    & 17.5\,TWh \\
     & & Electricity demand for hot water     & 1900 kWh/dwelling (2.7\,TWh) & 1900 kWh/dwelling (2.4\,TWh) \\
     & & Total heated floor area         & 176 million\,m$^\text{2}$ (122\,m$^\text{2}$ per dwelling) & 192 million\,m$^\text{2}$ (152\,m$^\text{2}$ per dwelling) \\
     & & Building properties of          & \multirow{2}{*}{Presented in \cref{tab:SFDSE} \cite{SEA2020}} & Presented in \cite{Mata2013}, \\
     & & \;\;\;\;construction materials  & & \;\;\;\;based on \cite{Boverket2011} \\
     & & Effective heat capacity         & varying           & fixed, 130,000 $\tfrac{\text{J}}{\text{K m}^\text{2}}$ \citep{Mata2013,Boverket2011} \\
     & & Indoor temperature setpoint (deadband) & $\theta^\text{set}$=19...23\,$^\text{o}$C, ($\theta^\text{set}\pm\,1^\text{o}$C)& 21.2\,$^\text{o}$C, (21.2\,...\,24\,$^\text{o}$C) \\
     & & Type and efficiency of electric heating & Presented in \cref{tab:Heatingshare} based on \cite{Nilsson2019,Boverket2011} & Presented in \cite{Mata2013}, based on \cite{Boverket2011}\\
     & & Power rating of the heating equipment & Presented in \cref{tab:ClimateZones}, based on \cite{Boverket2012} with \cite{Tangix} & Done in \cite{Nyholm2016}, based on \cite{Boverket2012}\\
     & & Temperature data                & Taken from \cite{Meteo2020}     & Taken from \cite{ECMWF2014}\\
     & & Solar irradiation               & -                               & Taken from \cite{ECMWF2014}\\
     & & Aggregate electric power        & 6.4\,GW (out of 12.9\,GW installed capacity)       & 7.3\,GW \\     
     & & Aggregate electric energy capacity & mean: 14.1\,GWh (7.0\,$\tfrac{\text{GWh}}{^\text{o}\text{C}}$), max: 21.3\,GWh & 19.3\,GWh (6.9\,$\tfrac{\text{GWh}}{^\text{o}\text{C}}$)\\
    \hline
    \multirow{4}{*}{ \rotatebox{90}{ \textbf{Power}} } & 
    \multirow{4}{*}{ \rotatebox{90}{ \textbf{system}} } & 
         Day-ahead energy market spot prices & Taken from Ref. \cite{NordPool} & Taken from Ref. \cite{NordPool} \\
     & & Imbalance settlement prices     & Taken from Ref. \cite{NordPool}        & - \\
     & & Reserve prices (specifically FCR-N) & Taken from Ref. \cite{Mimer}           & - \\
     & & Hourly mean system frequency    & Taken from Ref. \cite{FingridFreq}     & - \\
    \hline
    \end{tabular}}%
  \label{tab:InputData}%
\end{table}%
The input data, data sources, and parameter calculation methods are compiled in \cref{tab:InputData} and they are compared to a previous study in Sweden performed by \cite{Nyholm2016} with similar building stock models. The key features to be compared are power and energy capacity. Compared to an aggregate power of $7.3$\,GW in \cite{Nyholm2016}, we obtain a more conservative estimate of $6.4$\,GW which may be due to the fact that we do not model the auxiliary electric heater that is used as a backup in most heat pumps. Compared to the aggregate energy capacity of $19.3$\,GWh in \cite{Nyholm2016}, we obtain $14.1$\,GWh with our method which is significantly lower. However, \cite{Nyholm2016} assumes a wider indoor temperature deadband, i.e., $2.8^\text{o}$C compared to $2.0^\text{o}$C in our data. Therefore, we scale the energy capacity with respect to the indoor temperature deadband. The scaled energy capacity in \cite{Nyholm2016} (7.0\,$\tfrac{\text{GWh}}{^\text{o}\text{C}}$) is almost identical to the one obtained with our data (6.9\,$\tfrac{\text{GWh}}{^\text{o}\text{C}}$). 
As an intermediary result, we argue that our data gives a conservative estimate of the flexibility potential that is available in the Swedish SFD building stock.

\section{Numerical Investigation} \label{sec:Case}
The data from \cref{sec:Data} are used in model \cref{eq:StoAllT} to simulate the DA self-scheduling of a risk-averse aggregator of electric heating and heat pumps for Swedish SFDs. Since we are interested in the technical flexibility potential, we assume for simplicity that the aggregator is a price-taker and has no impact on the market clearing. The heating season is assumed to last from October 1$^\text{st}$ to May 31$^\text{st}$. 
\cref{fig:DataSEall} illustrates the long term hourly average ambient temperatures in the four areas (bottom) and the hourly mean prices (top).
We use 25 scenarios to represent the uncertainty from RT prices and frequency realizations. The same approach as in \cite{Herre2020TCL} is used to model uncertainty distributions of the thermal energy storage bounds.

\subsection{Time Series}
\begin{figure}[t]
\centering
\includegraphics[width=\linewidth,trim={0.2cm 0.65cm 0.2cm 0.2cm},clip]{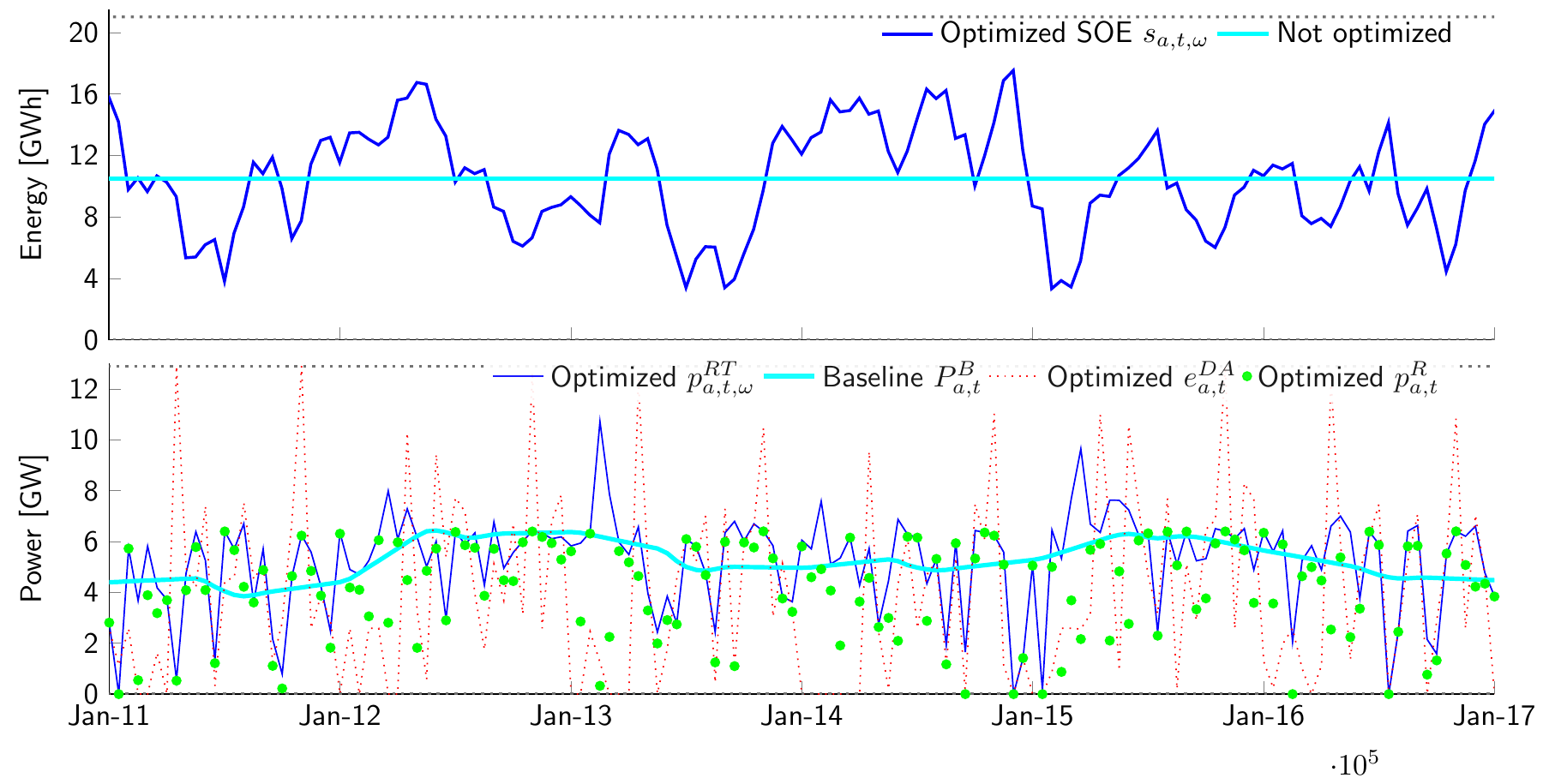}
\caption{Time series of summed variables over all areas. Top: expected value of state of energy over all scenarios. Bottom: RT power consumption (blue), baseline power (light blue), reserve bids (green $\bullet$), and DA energy bids (red dotted) in GWh/h. RT power consumption is given as the expected value over all scenarios.}
\label{fig:DataSEall}
\end{figure} 
\begin{figure}[t]
\centering
\includegraphics[width=\linewidth,trim={0.2cm 0.65cm 0.2cm 0.2cm},clip]{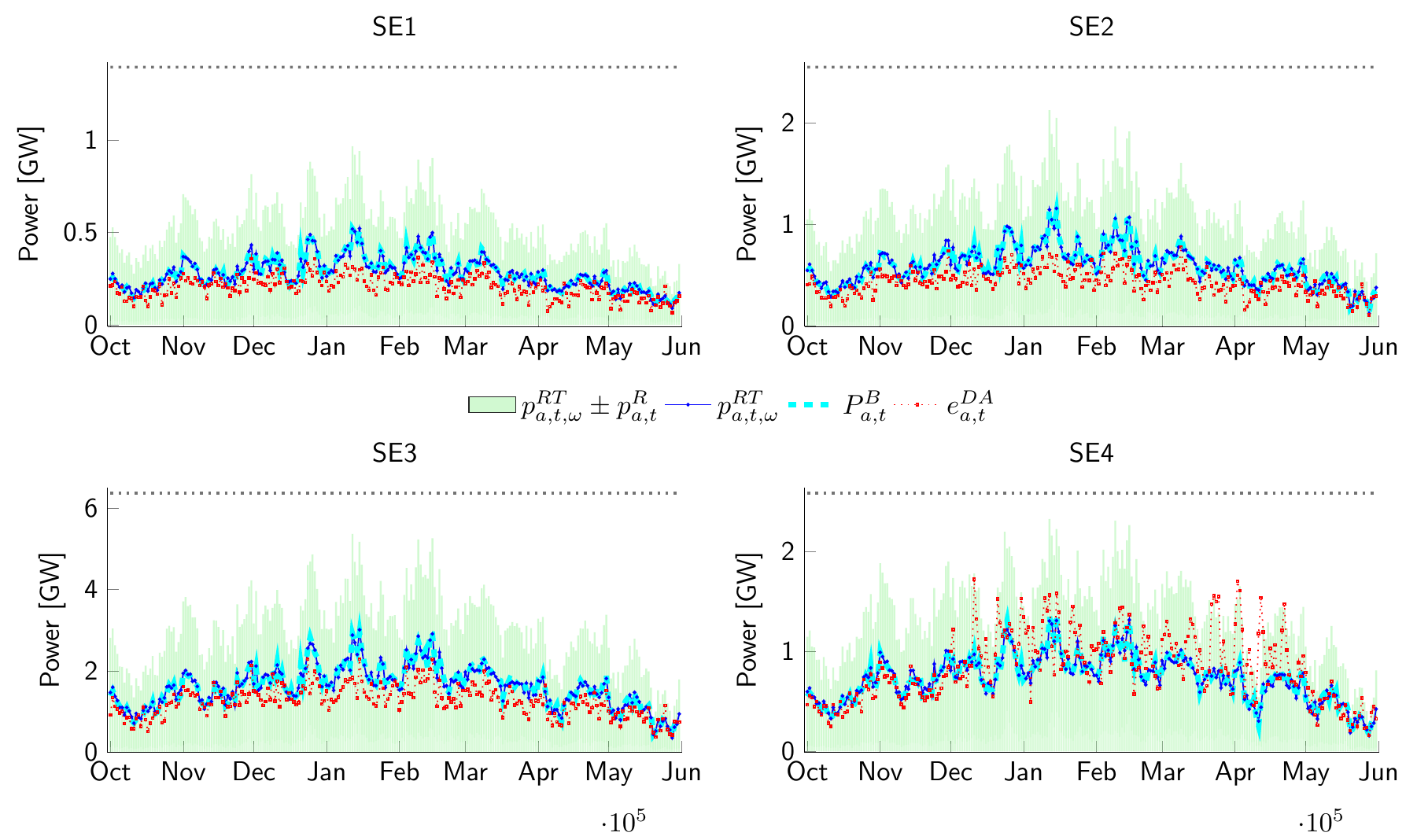}
\caption{Time series of individual areas as daily mean values of RT power consumption (blue), baseline power (light blue), reserve bids (green bars), and DA energy bids (red dotted). The RT power consumption is given as the expected value over all scenarios. The reserve bids are illustrated as a range ($\pm p^R_{a,t}$) around the expected RT consumption.}
\label{fig:DataSE1234daily}
\end{figure} 
\cref{fig:DataSEall} shows the time series of the decision variables that are summed over all price areas. The top figure shows the expected value of the aggregate state of energy of the thermal energy storage. The bottom figure shows the RT power consumption (blue), baseline power (light blue), reserve bids (green $\bullet$), and DA energy bids (red dotted) in GWh/h. It can be observed that energy arbitrage is exploited frequently between the DA energy market and the RT market. It can further be observed that reserve bids are generally higher when the baseline consumption is high. Reserve bids are mainly concentrated around half the maximum power ($\tfrac{\overline{\mathcal{P}}}{2}$) and 0, due to the symmetry requirement of the reserve.

\cref{fig:DataSE1234daily} shows the time series of the decision variables in each price area as daily mean values. 
It can be seen that the expected value of the real-time consumption closely follows the baseline power. While the DA energy bid is close to the RT consumption on the daily average, several arbitrage instances are visible. 
With respect to the direction of the imbalance between DA and RT markets, an interesting property can be observed in area SE4 compared to the remaining areas. While in SE1 to SE3, the RT consumption is largely higher than the DA bid, in SE4, the RT consumption is mostly lower than the DA bid.

\subsection{Sensitivity Analysis}
\begin{figure}[t]
  \centering 
  \includegraphics[width=0.85\textwidth,trim={0.2cm 0.65cm 0.2cm 0.2cm},clip]{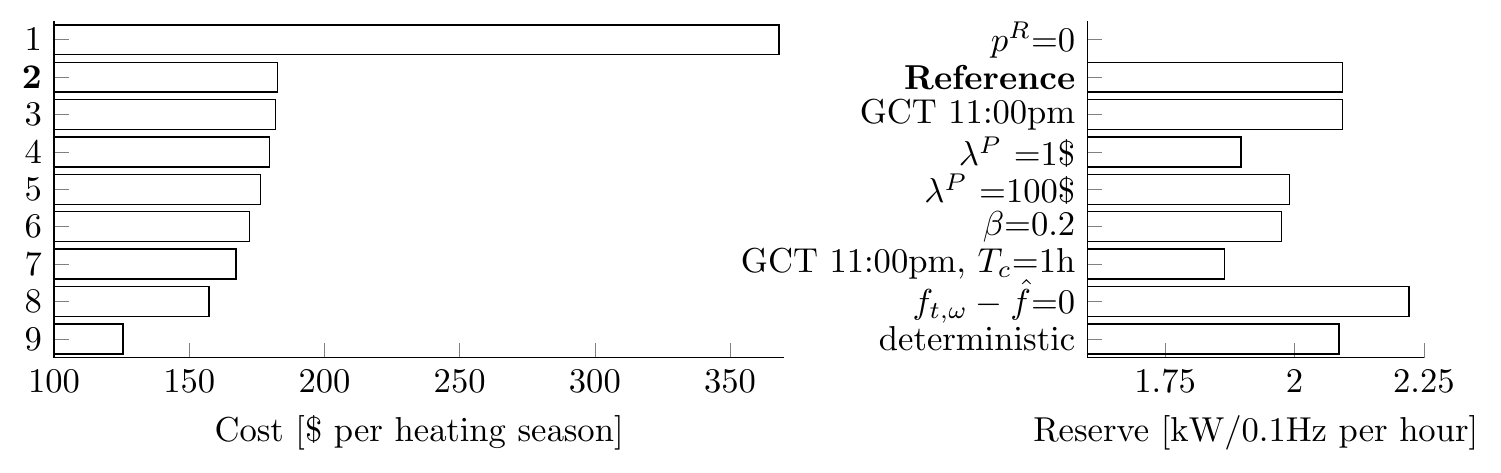}
  \caption{Comparison of cases 1 to 9 in terms of expected cost per SFD and mean hourly reserve per SFD during one heating season (October to May).}
  \label{fig:Comparison}
\end{figure}

\begin{table}[t]
  \centering
  \caption{Summary of case studies for the sensitivity analysis}
    \setlength{\tabcolsep}{4pt}
    \begin{tabular}{rl lll}
    \hline
      & Case                  &  change with respect to reference  \\
    \hline 
    1 & Energy arbitrage only           & $\lambda^R_t = 0 \; \forall t\in\mathcal{T}$  \\
    \textbf{2} & \textbf{Energy arbitrage \& reserve optimization} & \textbf{Reference} \\
    3 & Later gate closure time (GCT)   & GCT 11:00pm: $T_{lead}\text{=1h}$  \\
    4 & Lower SOE penalty               & $\lambda^P = 1$\textdollar  \\
    5 & Higher SOE penalty              & $\lambda^P = 100$\textdollar  \\
    6 & More risk-taking aggregator     & $\beta = 0.2$  \\
    7 & Later GCT and shorter contract period & $T_{lead}\text{=1h}, \;T_{c}\text{=1h}$ \\
    8 & Zero-mean guarantee for activation signal & $f_{t,\omega} -\hat{f}=0 \;\forall t \in \mathcal{T}, \forall \omega \in \Omega$ \\
    9 & Deterministic: perfect foresight$^{*}$  & $\mathbf{Y}' = \underset{\omega \in \Omega}{\mathbb{E}}[\mathbf{Y}_{\omega}]$\\
    \hline 
    \multicolumn{3}{l}{$^{*}$: \small{$\mathbf{Y}'$ refers to the deterministic parameters and $\mathbf{Y}$ are the stochastic parameters.}}
    \end{tabular} 
  \label{tab:sensitivity}%
\end{table}%
The reference case comprised 25 scenarios with risk- aversion parameter $\beta=0.5$ (in today's market setup with gate closure time (GCT) at 12:00\,pm on the day ahead). The SOE penalty is copmputed dynamically and is equal to the DA energy price in the last market interval of the optimization horizon, i.e., $\lambda^P=\tfrac{1}{N_a}\sum_{a\in\mathcal{A}}\lambda^{DA}_{\tau_h,a}$.
In addition to the reference case we investigate eight more cases by modifying one or more input parameters as summarized in \cref{tab:sensitivity}. 
Case 1 models the optimization of energy arbitrage, without any reserve bids ($p^R=0$). Case 2 is the reference case while case 3 models a delayed GCT at 11:00\,pm, 1\,hour before operation.
Cases 4 and 5 analyze different SOE penalties, $1$\,\textdollar \,and $100$\,\textdollar, respectively. Case 6 shows the results assuming that the aggregator was more risk-taking ($\beta=0.2$). Case 7 models a market that closes 1\,hour before each operating hour, similar to the current intraday energy market in Sweden. Case 8 assumes that the hourly energy activation of reserve bids has a zero mean character within each hour ($f_{t,\omega}-\hat{f}=0$), as assumed in \cite{Herre2020TCL}. The deterministic case with one scenario is modeled in case 9.
\cref{fig:Comparison} shows the reserve capacity and expected profit, here shown as expected cost per SFD, in each of the cases 1 to 9. The cases are sorted in descending order according to their expected cost per average SFD.

By comparing cases 3 and 7 with case 2, the benefit of shorter market timing is visible. Delaying the GCT (case 3) positively impacts the expected cost and reserve capacity. However, a shorter contract period (case 7) reduces the cost while reducing the reserve capacity as well. This is because the energy arbitrage is exploited more heavily with short market timing. 
By comparing cases 4 and 5 with case 2, the strong impact of the SOE penalty is illustrated. While different penalties can reduce the expected cost, they both negatively impact the available reserve capacity.
With more risk-taking behavior (case 6), both the expected cost and reserve capacity are reduced.
The guarantee of a zero-men activation signal (case 8) has the highest positive impact on both; expected cost and reserve capacity.

\section{Conclusions and Policy Implications} \label{sec:Conclusion}
In this paper, we model the share of the Swedish single family dwelling stock that is electrically heated as `aggregate thermal energy storage' in its respective price area. We use market and temperature data to model a risk-averse aggregator that bids the flexibility of the energy storage into energy and reserve markets as a two-stage chance-constrained problem and recast the problem as a linear program. Case studies show that each dwelling can contribute on average 2.1\,$\tfrac{\text{kW}}{\text{0.1Hz}}$ per hour during the heating season. We analyze different measures that may improve the economic feasibility or the available reserve capacity. 

From the results of the sensitivity analysis two main policy implications become apparent; they are based on case 7 discussed below in \cref{sec:Pol1}, and case 8 discussed in \cref{sec:Pol2}. Both case 7 and case 8 are discussed with respect to the reference case. 
\subsection{Market Timing: Gate Closure \& Contract Period} \label{sec:Pol1}
Obviously, from the perspective of the system operator that needs to carefully schedule reliable operation and reserve capacity in advance, a long lead time is favorable. Long lead times allow the operator to run contingency analysis and potentially procure or prepare capacity for extreme events. In Sweden, the last gate closure for FCR is currently at 6:00\,pm, where all 24 hourly bids ($T_c$=24\,h) for the next day must be submitted. 

We argue that a later gate closure has the following benefits. 
\begin{itemize}
    \item The business case for the TCL aggregator is enhanced since it can reduce the operational cost, as shown in this paper, \cref{fig:Comparison}.
    \item It can increase the amount of reserve bids from aggregators, as shown in \cite{Herre2020TCL}. Here, this effect is too small to be visible. This is partly because the modelling approach of aggregating the capacity of an entire price area reduces the impact from aggregate power and energy bounds.
    \item Any uncertainty related to power system operation decreases with reduced lead time. Further examples are outages and forecast errors of renewable energy.  
\end{itemize}
However, we acknowledge that gate closures have been designed with long lead times for a reason. This reason is mainly based on the assumption of a fossil fuel based generation mix that depends strongly on ramping constraints of thermal generators. Therefore, in most power system, a delayed gate closure would come at the cost of suboptimal thermal generation dispatch, as of today. 
On the other hand, as we approach higher shares of renewable power generation and more active demand response, the design focus on thermal generation might be challenged.

The exact time for an optimal gate closure can be obtained by a careful cost and benefit analysis for all involved actors and relies heavily on the generation mix and state-of-the-art of demand response rollout. We briefly lay out a sketch of the optimization model that would quantify such a system optimal gate closure:

A bilevel stochastic and chance-constrained optimization model can capture the interaction of TCL aggregator and system operator as a Stackelberg game. The system operator can be seen as the leader in the upper-level problem, with a fixed generation mix at its disposal. The TCL aggregator's problem can be easily integrated in the lower-level problem since it is convex (more specifically, linear). We can then solve the problem for a finite set of different gate closures and compare the results in terms of social welfare. 

A similar analysis can be done for the optimal value of the contract period $T_c$. \cref{fig:Comparison} clearly shows that a later gate closure (11:00\,pm), together with shortened contract periods ($T_c$=1\,h) is likely to significantly decrease the operating cost of the TCL aggregation. However, this happens at the disadvantage of decreased reserve capacity. 

\subsection{Guarantees for the Activation Signal} \label{sec:Pol2}
For primary reserves (FCR), in Europe, the reserve activation signal is the power system frequency which can be based on local measurements. Frequency is highly stochastic in nature. Over large periods of time, however, its mean value is approximately equal to the nominal frequency. 

The reserve provision from TCLs is limited by the remaining state of energy in the aggregate thermal energy storage. When optimizing reserve bids, a TCL aggregator must consider all possible frequency activation trajectories and ensure sufficient energy levels. This is mainly because the activated energy is directly related to mean frequency deviation in that market interval. 

If the system operator was to establish guarantees for the activation signal, this may significantly help TCL aggregators to provide more reserve capacity and increase their revenue (or reduce their operational cost). Such guarantees could, for instance, entail the splitting of the frequency into two separate signals where one has a zero-mean character and the other corrects for offsets on a longer time scale. Zero-mean characters have been introduced in some North American Independent System Operators such as California Independent System Operator's Regulation Energy Management \cite{CaliforniaISO2011}. 

In Europe, however, zero-mean guarantees do not yet exist. The origin of the design space of reserve markets lies in the guidelines of the European Network of Transmission System Operators for electricity, ENTSO-e. This body specifies the parameters of reserve markets within which the national system operators can then design their own market. As of today, there is still no move towards zero-mean guarantees which constitutes a major hindrance for operators of energy limited resources, such as aggregated demand response discussed in this paper.

\newpage
\section*{Appendix}
\begin{figure}[t]
\centering
\includegraphics[width= \columnwidth]{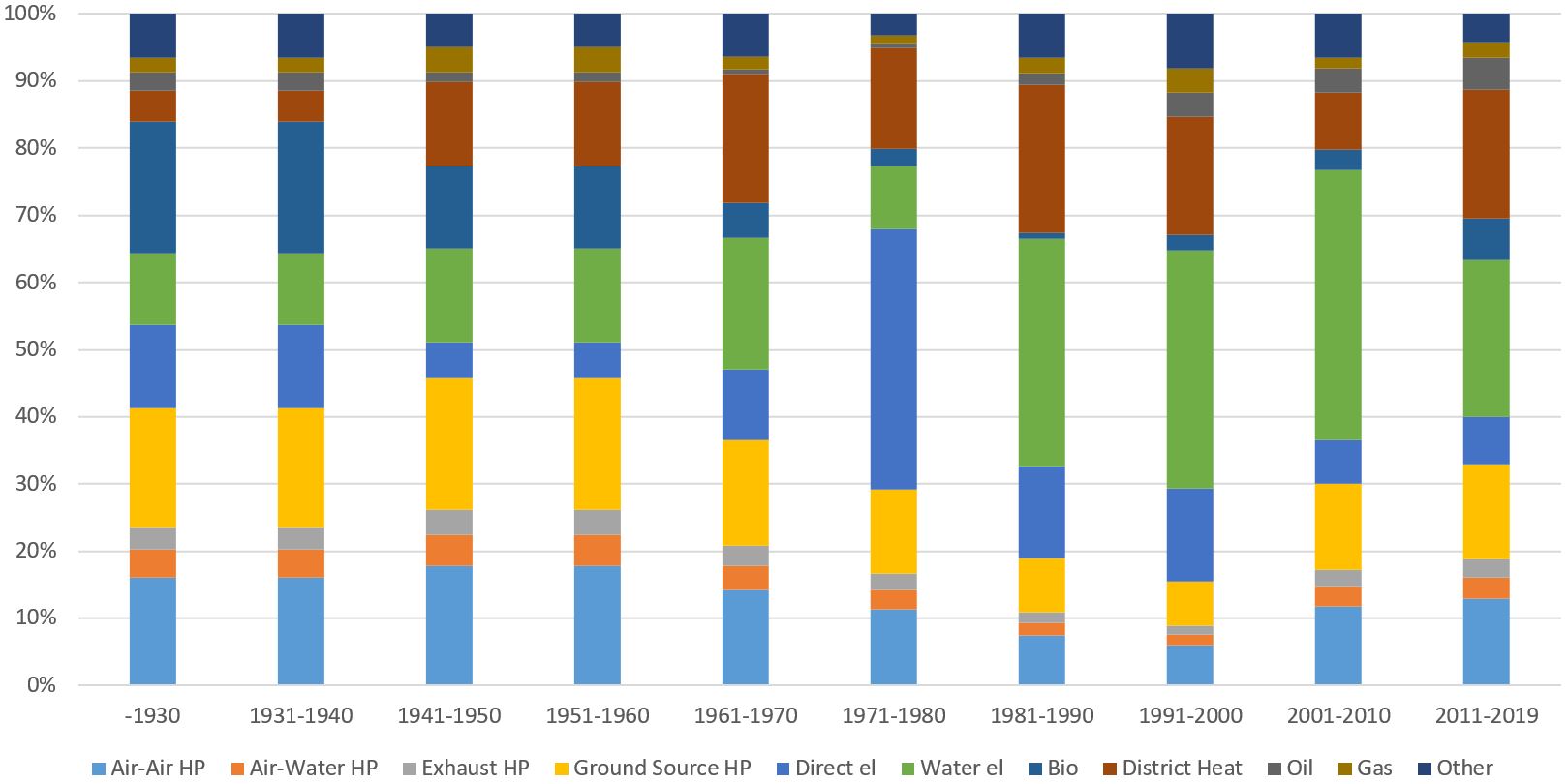}
\caption{Share of installed heating system in Swedish SFDs by decade}
\label{fig:HeatingShareX}
\end{figure}

\cref{fig:HeatingShareX} shows the share heating type that was installed in Sweden per decade.


\bibliographystyle{cas-model2-names}
\bibliography{mainbib}

\end{document}